\documentclass[aps,pra,superscriptaddress,twocolumn,showpacs]{revtex4}
\usepackage{graphicx}
\usepackage{amssymb}
\usepackage{amsmath}
\usepackage{bm}

\usepackage{tikz}
\usetikzlibrary{shapes, positioning}

\usepackage{color}

\begin{document}
\setcounter{page}{0}

\title[]{Hidden quantum-classical correspondence in chaotic billiards revealed by mutual information}
\author{Kyu-Won \surname{Park}}
\email{parkkw7777@gmail.com}
\affiliation{Department of Mathematics and Research Institute for Basic Sciences, Kyung Hee University, Seoul, 02447, Korea}

\author{Soojoon \surname{Lee}}
\affiliation{Department of Mathematics and Research Institute for Basic Sciences, Kyung Hee University, Seoul, 02447, Korea}
\affiliation{School of Computational Sciences, Korea Institute for Advanced Study, Seoul 02455, Korea}

\author{Kabgyun \surname{Jeong}}
\email{kgjeong6@snu.ac.kr}
\affiliation{Research Institute of Mathematics, Seoul National University, Seoul 08826, Korea}
\affiliation{School of Computational Sciences, Korea Institute for Advanced Study, Seoul 02455, Korea}

\date{\today}

\begin{abstract}
Avoided level crossings, commonly associated with quantum chaos, are typically interpreted as signatures of eigenstate hybridization and spatial delocalization, often viewed as ergodic spreading. We show that, contrary to this expectation, increasing chaos in quantum billiards enhances mutual information between conjugate phase space variables, revealing nontrivial correlations. Using an information-theoretic decomposition of eigenstate entropy, we demonstrate that spatial delocalization may coincide with increased mutual information between position and momentum. These correlations track classical invariant structures in phase space and persist beyond the semiclassical regime, suggesting a robust information-theoretic manifestation of quantum-classical correspondence.
\end{abstract}

\maketitle

\section{Introduction}
Avoided crossings (A.C.) between quantum energy levels are widely introduced in textbooks as hallmark signatures of level interaction~\cite{vNW29,Landau32,Zener32,StockmannBook99,Moise2022book}. When two energy levels approach each other under a parametric variation of the Hamiltonian, their coupling induces level repulsion, preventing a true degeneracy. The phenomenon of avoided crossings manifests across a wide range of physical systems, including atomic~\cite{Chilcott21, Migliore24}, mesoscopic~\cite{GarciaMarch15}, and even astrophysical contexts such as gravitational wave spectra~\cite{Sotani20}. A particularly intriguing feature of interacting eigenstates in this context is their ability to hybridize and reorganize under level proximity, leading to significant changes in spatial or phase-space structure~\cite{BerryTabor77,BGS84}.

A particularly convenient platform for probing such avoided crossings is provided by chaotic quantum billiards, where a point particle undergoes elastic reflections within a two-dimensional enclosure~\cite{StockmannBook99, BerryRobnik84}. Depending on the boundary geometry, the corresponding classical dynamics can be either integrable or chaotic, with even modest boundary deformations typically giving rise to eigenstate hybridization and a proliferation of avoided crossings~\cite{StockmannBook99, Graef92,Ellegaard95}.

To characterize the spectral behavior of chaotic systems, random matrix theory (RMT) has long served as a foundational framework~\cite{StockmannBook99, Izrailev90}. RMT captures universal properties of spectral fluctuations and level repulsion, but it offers no direct access to the structure of the eigenstates themselves. In response, various measures of eigenstate delocalization such as Shannon entropy and participation ratio have been developed to quantify the extent of wavefunction spreading in configuration or momentum space~\cite{Park18,Park22,Prasad24,Hu24}. While these global metrics capture overall broadening, they cannot distinguish between random dispersion and structured internal correlations. As a result, existing approaches may overlook hidden structure that persists even in strongly chaotic regimes.

In this work, we address this conceptual gap by applying an \emph{information-theoretic decomposition} of the entropy of a quantum state~\cite{CoverThomas99, Groisman05, Modi12}. Specifically, we decompose the total entropy of the phase-space distribution into conditional components and mutual information (MI), enabling a quantification of internal structure within the eigenstate that goes beyond global delocalization. Our study focuses on chaotic billiards with quadrupole and oval geometries, where classical instability is tunable and avoided crossings are abundant. By analyzing both configuration-space and phase-space distributions, we demonstrate that increasing chaos not only leads to spatial delocalization but also enhances mutual information, revealing nontrivial internal structure in the phase-space representation of the eigenstate.

This result challenges the conventional view that chaos suppresses structure, and instead suggests that partial disorder can amplify internal correlations or uncover hidden organization. Our information-theoretic framework captures this phenomenon beyond the semiclassical regime~\cite{Takami92,MainieroPorter2007}, offering a refined and robust form of quantum-classical correspondence.

Moreover, because our analysis relies solely on the probability distributions of eigenstates, the proposed method naturally extends to various wave-based or open systems, including those with leaky boundaries or non-Hermitian effects~\cite{Weisbuch92, Song13, Shin16, Zhu10}. This suggests that the observed phenomenon is not exclusive to closed billiards.

\emph{Roadmap.}---In Sec.~II, we outline the conceptual and computational framework. Section~III describes informational behavior in configuration and phase space for the quadrupole billiard. In Sec.~IV, we extend our results to oval billiards. Section~V discusses classical invariant tori and multi-branch functionality. Finally, Sec.~VI summarizes our key findings and discusses their broader implications for wave-based phenomena and open chaotic systems.

\section{Conceptual and Computational Framework}
\subsection{Information-Theoretic Decomposition}

To analyze wavefunction behavior near avoided crossings, it is essential to distinguish between different forms of uncertainty and correlation embedded in a quantum system. While the Shannon entropy quantifies the total uncertainty, it does not reveal how this uncertainty is partitioned between variables, or how much information is shared between them.

For two random variables \(X\) and \(Y\), the joint entropy can be decomposed as
\begin{align}
H(X,Y) = H(X \mid Y) + H(Y \mid X) + I(X;Y),
\end{align}
where:
\begin{itemize}
    \item \(H(X \mid Y)\) is the conditional entropy of \(X\) given \(Y\), representing the residual uncertainty in \(X\) after knowing \(Y\);
    \item \(H(Y \mid X)\) is the conditional entropy of \(Y\) given \(X\);
    \item \(I(X;Y)\) is the mutual information, quantifying the amount of information shared between \(X\) and \(Y$)~\cite{CoverThomas99}.
\end{itemize}

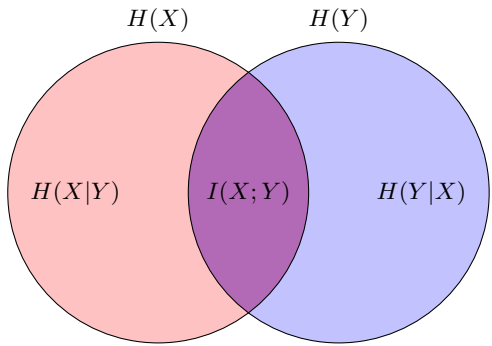
\begin{figure}[htbp]
\centering
\begin{tikzpicture}
\path[fill=red!30, opacity=0.8] (-1.2,0) circle (2cm);
\path[fill=blue!30, opacity=0.8] (1.2,0) circle (2cm);
\begin{scope}
    \clip (-1.2,0) circle (2cm);
    \fill[violet!60, opacity=0.95] (1.2,0) circle (2cm);
\end{scope}
\draw (-1.2,0) circle (2cm);
\draw (1.2,0) circle (2cm);
\node at (-2.3,0) {\small $H(X|Y)$};
\node at (2.3,0) {\small $H(Y|X)$};
\node at (0,0) {\small $I(X;Y)$};
\node at (-1.2,2.3) {\small $H(X)$};
\node at (1.2,2.3) {\small $H(Y)$};
\end{tikzpicture}
\caption{Venn diagram illustrating the decomposition of joint entropy \(H(X,Y)\) into conditional entropies and mutual information.}
\label{fig:venn-entropy}
\end{figure}

This decomposition is central to our analysis. While \(H(X,Y)\) captures the total uncertainty of a quantum state, its components help reveal whether entropy increases stem from incoherent delocalization (via conditional entropies) or from coherent internal correlations (via mutual information). In the context of avoided crossings, this distinction helps clarify whether changes in the wavefunction arise from random spreading or from structured reorganization of internal degrees of freedom.

Crucially, this decomposition enables a direct comparison between different representations such as configuration space and phase space on equal footing, based on how each encodes uncertainty and correlation.

\subsection{Theoretical Framework and Physical System}
\begin{figure*}
\centering
\includegraphics[width=13.5cm]{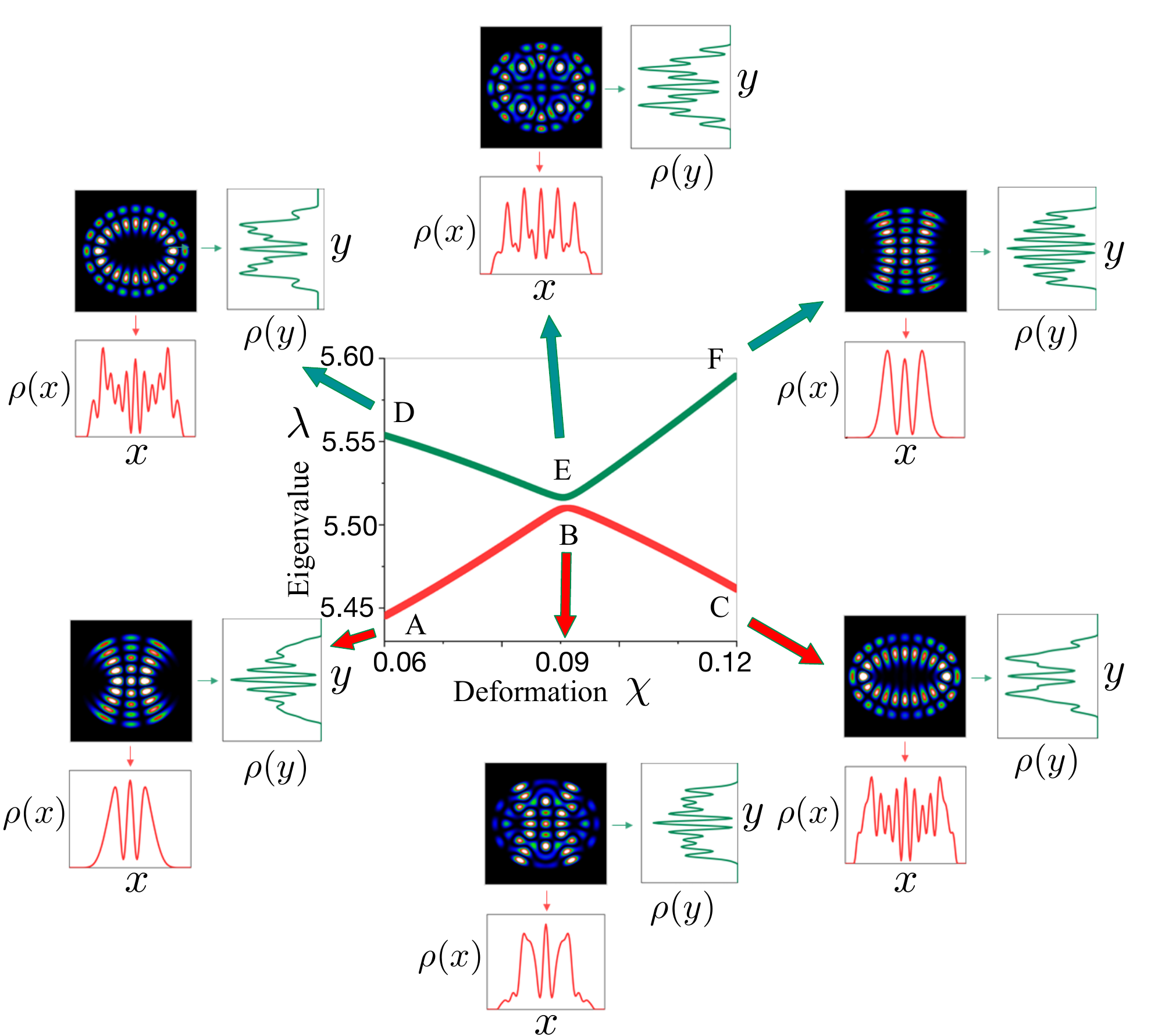}
\caption{Avoided crossing in a chaotic billiard as the deformation parameter $\chi$ varies. Eigenvalue trajectories of two interacting modes are shown as orange and green solid lines, with hybridization occurring near $\chi \approx 0.09$. Insets display spatial probability densities $\rho(x,y)$ and corresponding marginals $\rho(x)$ and $\rho(y)$ for six representative eigenstates: A--C (bottom row, \textbf{Mode~1}) and D--F (top row, \textbf{Mode~2}). Near the crossing center (states B and E), eigenfunctions broaden anisotropically while retaining directional asymmetry.}
\label{Figure-1}
\end{figure*}

We apply this information-theoretic framework to chaotic quantum billiards. Dynamical billiards describe the motion of a point particle confined to a bounded domain, moving along straight-line trajectories and undergoing specular reflection at the boundaries without energy loss. In the classical regime, this motion is governed by the Hamiltonian $\mathcal{H}(q, p) = \frac{p^2}{2m} + V(q)$, where the potential \(V(q)\) is zero inside the billiard and infinite at the boundary. In the quantum regime, classical trajectories are replaced by solutions to the time-independent Schr\"{o}dinger equation, $-\frac{\hbar^2}{2m} \nabla^2 \psi(q) = E \psi(q)$, which reduces to the Helmholtz equation, $(\nabla^2 + k^2)\psi(q) = 0$, under Dirichlet boundary conditions~\cite{StockmannBook99,Park22}.

We consider two chaotic billiard geometries: the quadrupole and the oval. The quadrupole billiard has a boundary defined in polar coordinates as
\begin{align}
\rho(\theta) = 1 + \chi \cos(2\theta)
\end{align}
~\cite{Park22}. The oval billiard, on the other hand, is generated by deforming an ellipse (\(\varepsilon = 0\)) in the \(x\)-direction, with a boundary defined by
\begin{align}
\frac{x^2}{a^2} + \bigl(1 + \varepsilon x\bigr)\frac{y^2}{b^2} = 1,
\end{align}
where \(\varepsilon\) is the deformation parameter~\cite{Park22a}. In this study, we fix \(a = 1.0\) and \(b = 1.03\). Both geometries explicitly break integrability and support fully chaotic classical and quantum dynamics. We solve the corresponding Helmholtz equation numerically with Dirichlet boundary conditions to obtain eigenvalues and eigenfunctions for our analysis.

\begin{figure}
\centering
\includegraphics[width=7.5cm]{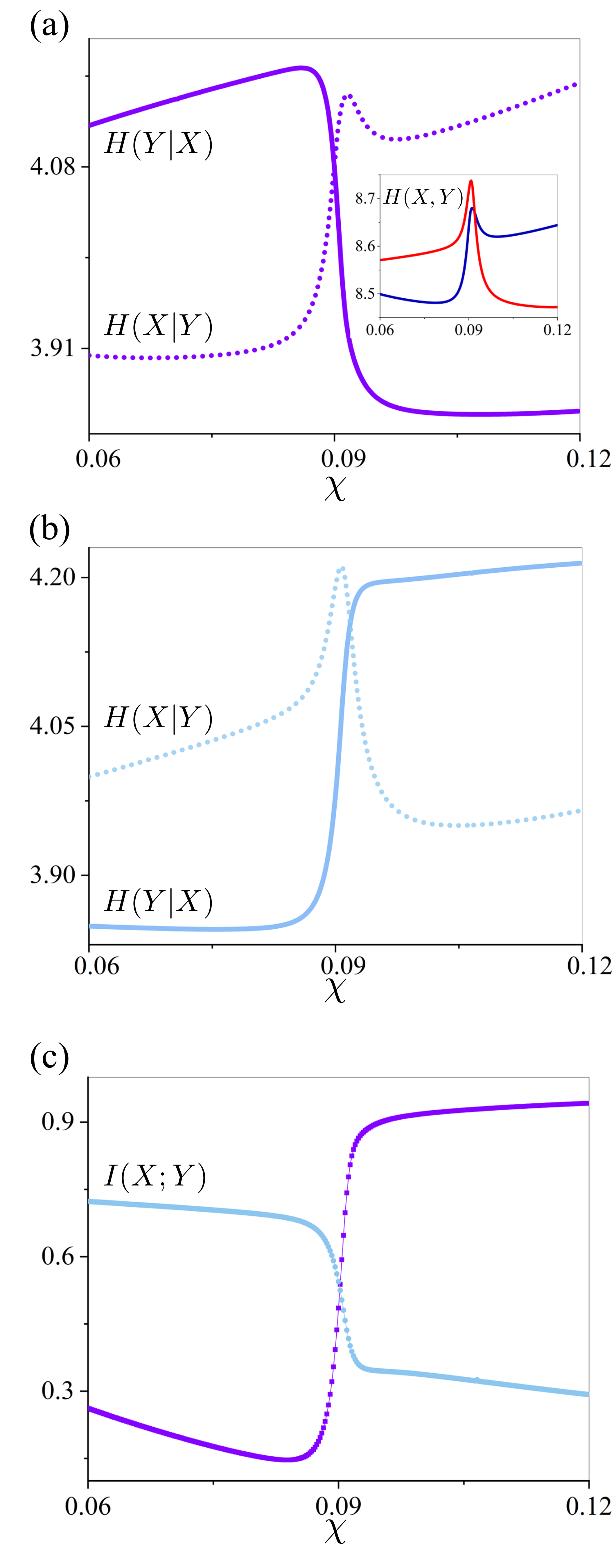}
\caption{Entropy decomposition across an avoided crossing in configuration space. (a) Conditional entropies for \textbf{Mode~1}: $H(Y|X)$ (solid purple) and $H(X|Y)$ (dotted purple). (b) Conditional entropies for \textbf{Mode~2}: $H(Y|X)$ (solid blue) and $H(X|Y)$ (dotted blue). (c) Mutual information $I(X;Y)$ for \textbf{Mode~1} (purple) and \textbf{Mode~2} (blue). Inset in (a): total entropy $H(X,Y)$. Results indicate directional delocalization without the emergence of structural correlation.}
\label{Figure-2}
\end{figure}

\subsection{Numerical Methods and Phase-Space Representation}

\begin{figure*}
\centering
\includegraphics[width=13.5cm]{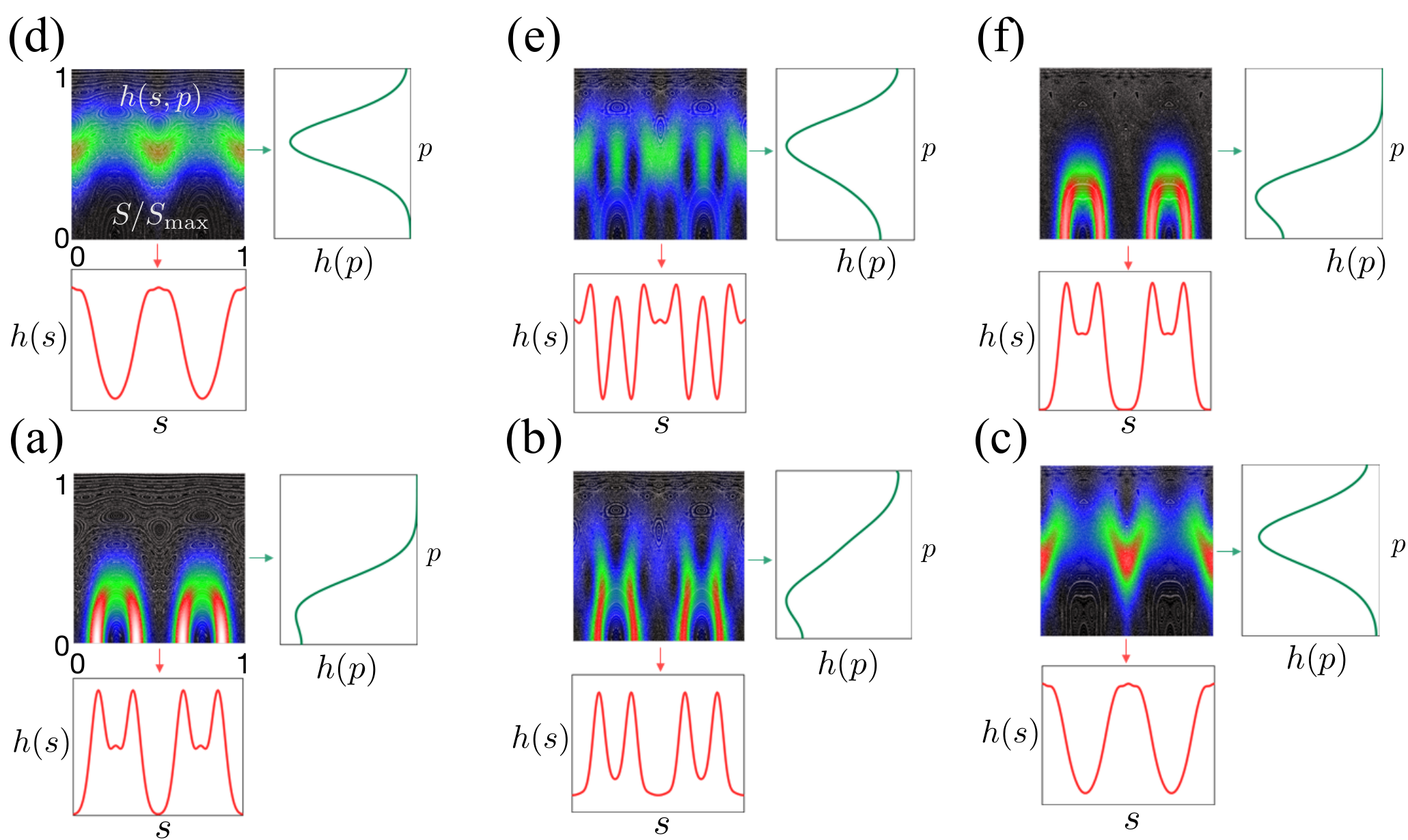}
\caption{Husimi distributions \(h(s, p)\) for six representative eigenmodes near the avoided crossing, with marginal densities \(h(s)\) and \(h(p)\) shown along the horizontal and vertical axes, respectively. Each subpanel shows a 2D Husimi projection (center), marginal momentum density \(h(p)\) (left), and marginal boundary projection \(h(s)\) (bottom). Top row: \textbf{Mode~2} (D--F); bottom row: \textbf{Mode~1} (A--C). As \(\chi\) increases, distributions become multi-lobed along \(s\) while remaining localized along \(p\), indicating directional hybridization in phase space.}
\label{Figure-3}
\end{figure*}

\begin{figure}
\centering
\includegraphics[width=7.0cm]{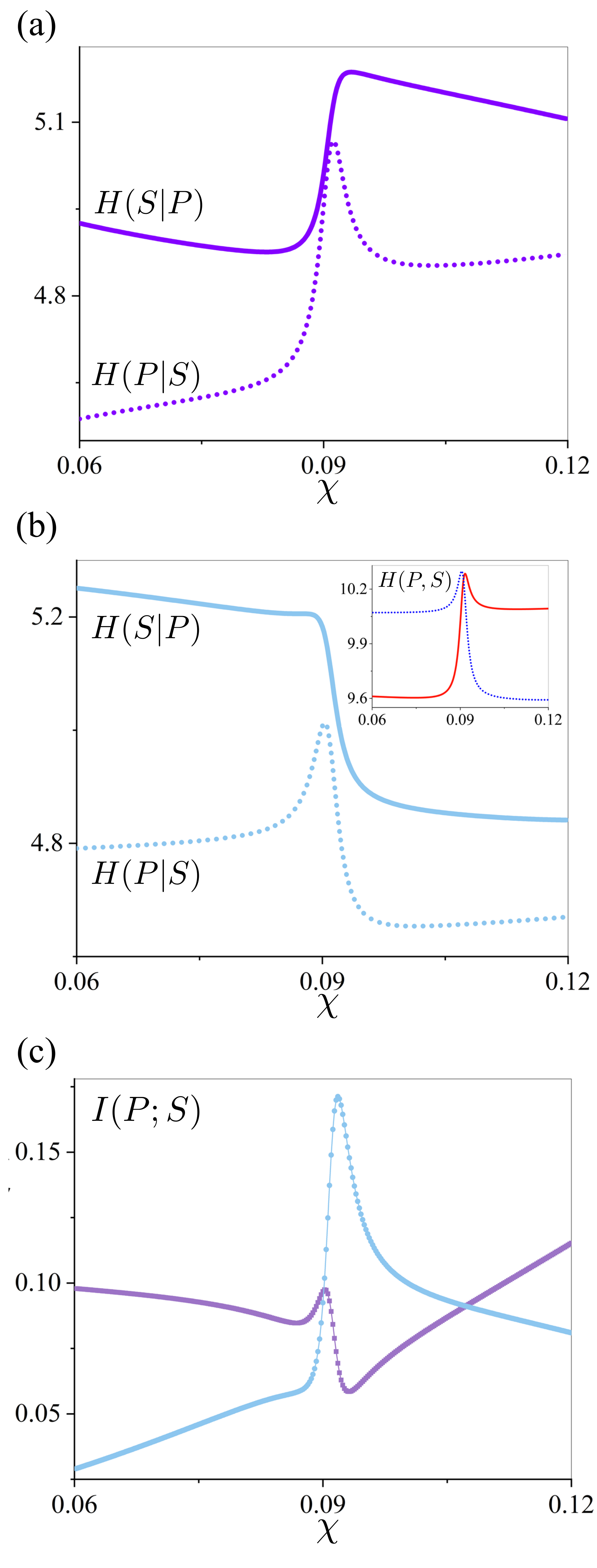}
\caption{Entropy decomposition in phase space across an avoided crossing. (a), (b): Conditional entropies \(H(S|P)\), \(H(P|S)\) for Modes 1 and 2. (c): Mutual information \(I(S;P)\) increases sharply in both cases but reaches significantly higher values for \textbf{Mode~1} (light blue with circles), indicating stronger internal correlation. Inset: total entropy \(H(S,P)\) peaks concurrently, revealing enhanced coherence alongside increasing delocalization.}
\label{Figure-4}
\end{figure}

\begin{figure*}[htbp]
\centering
\includegraphics[width=13.5cm]{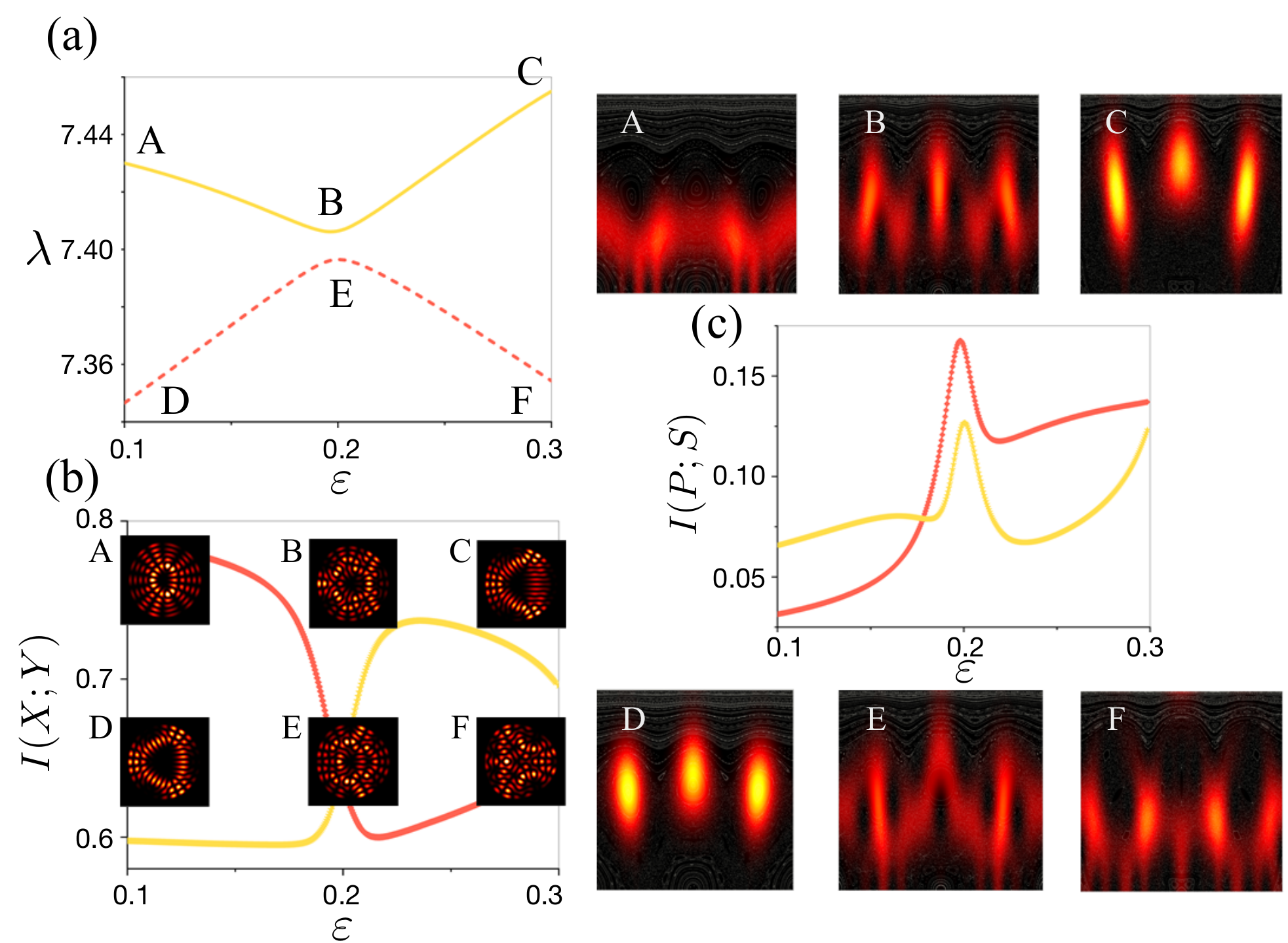}
\caption{(a) Avoided crossing in the oval billiard as a function of deformation \(\varepsilon\), showing eigenvalue trajectories for two branches (orange and red dashed lines). (b) Real-space probability distributions \(\rho(x,y)\) for sampled states, along with corresponding mutual information \(I(X;Y)\) shown as bar plots. (c) Phase-space mutual information \(I(S;P)\) for two eigenstate pairs; red and yellow solid lines indicate distinct hybridized modes. In both cases, \(I(S;P)\) peaks near the crossing, confirming that coherence enhancement is a robust feature across chaotic geometries.}
\label{Figure-5}
\end{figure*}

We use the boundary element method (BEM) to solve the Helmholtz equation
\begin{align}
(\nabla^2 + n^{2}k^2)\psi(\mathbf{r}) = 0
\end{align}
with Dirichlet boundary conditions (\(\psi = 0\) on the billiard wall) for closed microcavities~\cite{Wiersig02}. The deformation parameters \(\chi\) for the quadrupole billiard and \(\varepsilon\) for the oval billiard are varied to trace eigenvalue trajectories and locate avoided crossings. We assume transverse-magnetic (TM) polarization and set the refractive index to \(n = 3.3\), where \(k\) is the wave number and \(\psi\) is the \(z\)-component of the electric field.

\begin{figure}
\centering
\includegraphics[width=7.5cm]{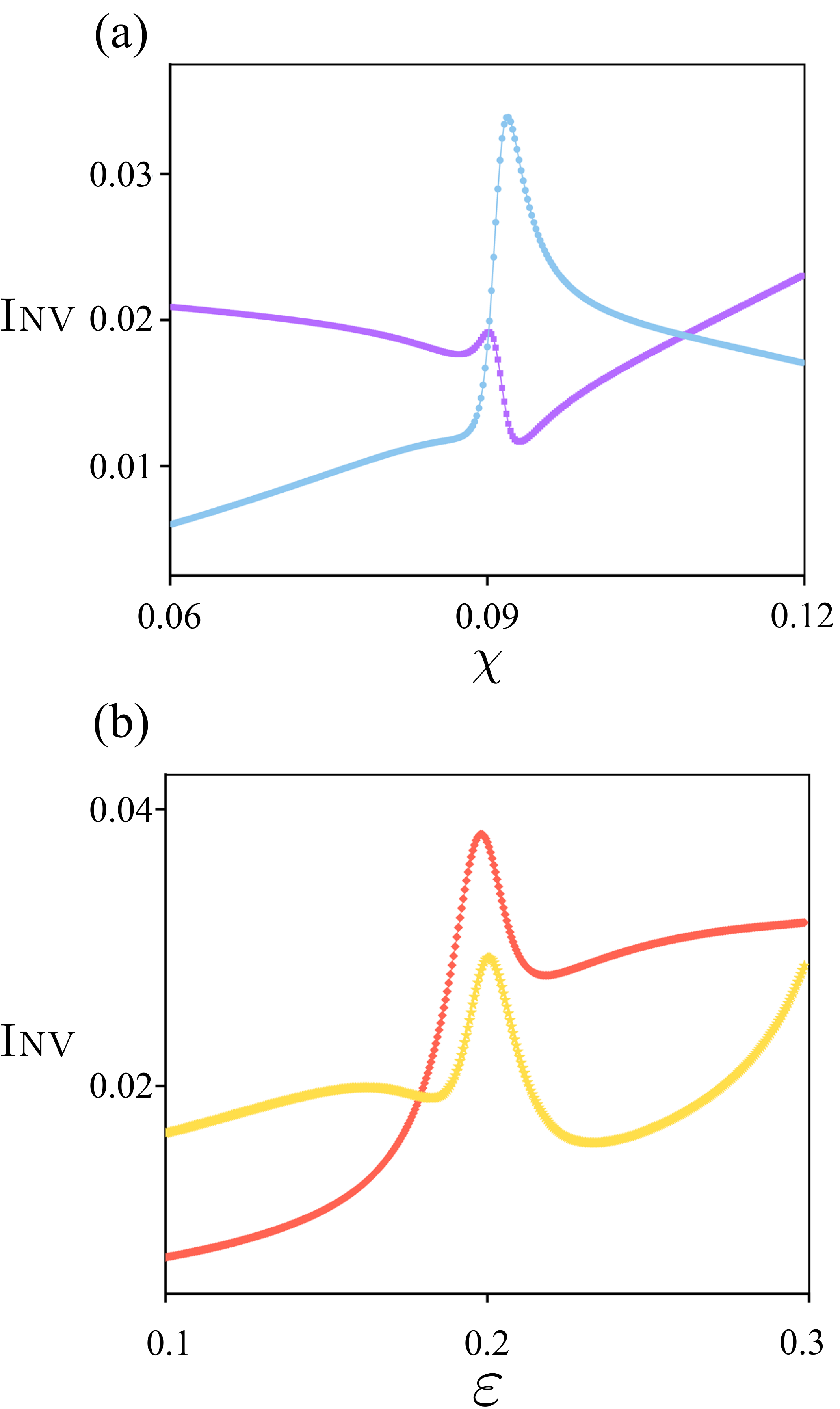}
\caption{Normalized mutual information, denoted \textsc{Inv}. (a), (b): \textsc{Inv} for quadrupole and oval billiards, respectively. Both exhibit a structural peak near the avoided crossing similar to that of \(I(S;P)\), but with smaller absolute values.}
\label{Figure-6}
\end{figure}

To analyze phase-space behavior, we compute the Husimi distribution \(h(s, p)\) for each eigenstate. Here, \(s\) denotes the normalized arc-length coordinate along the boundary, and \(p = \sin(\chi')\) is the conjugate momentum, with \(\chi'\) representing the angle of incidence. The Husimi distribution is constructed by projecting the boundary-normal derivative \(\partial_n \psi(s')\) onto a family of minimum-uncertainty coherent states~\cite{Park22a,Crespi93}:
\begin{align}
h(s, p) = \left|\int \partial_n \psi(s')\, \xi(s'; s, p)\, ds'\right|^2,
\end{align}
where \(\xi(s'; s, p)\) is a Gaussian window centered at \((s, p)\) in phase space.

From these distributions, we compute marginal and joint probabilities to evaluate the Shannon entropy, conditional entropies, and mutual information. Tracking these quantities as the deformation parameter varies across an avoided crossing enables us to disentangle how wavefunction uncertainty evolves due to spatial delocalization versus inter-variable correlation. This forms the basis for our comparative analysis between configuration and phase space.

\section{Informational Behavior in Configuration and Phase Space: Quadrupole Billiard}
\subsection{Configuration-Space Analysis}

We begin by analyzing how informational structure evolves in configuration space across an avoided crossing in the quadrupole billiard. The deformation parameter \(\chi\) is varied over the range \([0.06, 0.12]\) with a fine scanning step of \(\Delta\chi = 1 \times 10^{-5}\), allowing us to resolve subtle variations in eigenvalue behavior and associated eigenmodes.

For each eigenmode, we compute the joint probability density \(\rho(x,y) = |\psi(x,y)|^2\) on a uniform \(100 \times 100\) grid. From this, we derive the marginal distributions as \(\rho_X(x_i) = \sum_{j=1}^{100} \rho(x_i, y_j)\), \(\rho_Y(y_j) = \sum_{i=1}^{100} \rho(x_i, y_j)\), and compute the marginal entropies \(H(X) = -\sum_i \rho_X(x_i) \log \rho_X(x_i)\), \(H(Y) = -\sum_j \rho_Y(y_j) \log \rho_Y(y_j)\), and the joint entropy \(H(X,Y) = -\sum_{i,j} \rho(x_i, y_j) \log \rho(x_i, y_j)\). To ensure numerical stability, terms with \(\rho=0\) are regularized as \(\rho \log \rho \rightarrow 0\). Mutual information is then defined as \(I(X;Y) = H(X) + H(Y) - H(X,Y)\).

The joint entropy can also be decomposed as
\[
H(X,Y) = H(X|Y) + H(Y|X) + I(X;Y),
\]
allowing us to separate contributions from directional uncertainty and inter-variable correlation. This yields the conditional entropies \(H(Y|X) = H(Y) - I(X;Y)\) and \(H(X|Y) = H(X) - I(X;Y)\), which reflect residual uncertainty along one variable given knowledge of the other.

Figure~\ref{Figure-1} shows an avoided crossing between two eigenvalue branches, with representative eigenmodes labeled A--F. The orange branch (\textbf{Mode~1}) contains states A--C, while the green branch (\textbf{Mode~2}) contains D--F. The marginal densities \(\rho_X(x)\) and \(\rho_Y(y)\), shown as projections of each eigenmode, reveal directional delocalization along the \(x\)- or \(y\)-axis. Near the crossing center (\(\chi \approx 0.09\)), eigenmodes B and E exhibit strong hybridization and increased spatial extent. This delocalization, however, is anisotropic, with dominant spreading along a single coordinate axis ''indicating a shift in the dominant direction of uncertainty, rather than increased inter-variable coupling.

To quantify this behavior, we examine the entropy-based measures defined above. Figure~\ref{Figure-2} presents conditional entropies \(H(X|Y)\), \(H(Y|X)\), and mutual information \(I(X;Y)\) as functions of \(\chi\). In panel (a), \textbf{Mode~1} (orange) shows a peak in \(H(X|Y)\) near the crossing, while \(H(Y|X)\) remains relatively flat. Panel (b) shows the opposite trend for \textbf{Mode~2} (green). Mutual information in panel (c) varies smoothly in both cases, without a pronounced peak. This reversal in conditional entropy trends reflects a directional exchange in uncertainty: the dominant axis of delocalization switches across the avoided crossing. The entropy peak in configuration space thus appears to stem primarily from anisotropic spreading, rather than a marked enhancement of internal correlation.

Although \(I(X;Y)\) evolves continuously, its relatively low magnitude suggests that the joint distributions remain close to separable, with only modest inter-axis dependence. These results confirm that configuration-space entropy measures are sensitive to directionality in delocalization, but may not fully capture the emergence of coherent internal structure.

Notably, the observed coherent superposition in configuration space often corresponds to a breakdown of well-defined quantum numbers, rather than a reorganization aligned with classical dynamics~\cite{StockmannBook99,Borgonovi16, Prosen07, Castagnino07}. Such hybridized spatial patterns challenge straightforward classification by nodal structure or separable quantum labels. This highlights a key limitation of configuration-space analysis: it effectively detects anisotropy, but offers limited insight into structural correlations that may manifest more clearly in phase space. In the following section, we demonstrate that phase-space analysis reveals a richer picture of the internal organization underlying avoided crossings.

\subsection{Phase-Space Analysis}
We now examine the informational structure of eigenmodes in phase space, where the coordinates \(s\) and \(p\) form a canonical pair of conjugate variables. Unlike the spatial coordinates \((x, y)\) in configuration space, the pair \((s, p)\) is dynamically linked through Hamiltonian mechanics. These variables satisfy the canonical Poisson bracket relation,
\begin{align}
\{q_i, p_j\} = \delta_{ij},
\end{align}
which yields Hamilton's equations: \(\dot{q}_i = \partial \mathcal{H} / \partial p_i\), \(\dot{p}_i = -\partial \mathcal{H} / \partial q_i\). While these relationships determine classical trajectories, they do not necessarily imply statistical or informational dependence between the variables in a quantum state. For instance, whispering-gallery modes often yield Husimi distributions that are highly localized along one axis, resulting in near-zero mutual information. This illustrates that despite their canonical coupling, \(s\) and \(p\) may exhibit marginal statistical independence depending on the eigenstate structure.

Therefore, although \(s\) and \(p\) are dynamically conjugate, the degree of informational dependence between them must be evaluated from the structure of the quantum state itself. The Husimi distribution \(h(s, p)\) provides a smooth, positive-definite representation of this structure. In this framework, mutual information \(I(S;P)\) quantifies inter-variable statistical dependence, while conditional entropies \(H(S|P)\) and \(H(P|S)\) measure directional delocalization across the phase-space axes.

Following the computational procedure outlined in Sec.~2.3, we evaluate the Husimi distribution and compute its marginal and joint entropies. Although the methodology parallels that of configuration-space analysis, the interpretation differs fundamentally due to the canonical nature of \((s, p)\). Figure~\ref{Figure-3} presents Husimi distributions and their marginals for six representative eigenmodes across an avoided crossing. Near the crossing point (\(\chi \approx 0.09\)), the distributions exhibit marked asymmetry: the marginal density along \(s\) becomes multi-lobed, indicative of hybridization and structural complexity, while the marginal along \(p\) remains narrow and localized.

Figure~\ref{Figure-4} displays the entropy-based measures derived from these distributions. In both modes, the conditional entropy \(H(P|S)\) shows a pronounced peak near the avoided crossing, whereas \(H(S|P)\) remains relatively flat and consistently higher in magnitude. This asymmetry suggests that knowledge of \(s\) offers greater predictive power about \(p\) than vice versa. Concurrently, the mutual information \(I(S;P)\) also peaks sharply at the crossing, revealing an increase in coherence between the conjugate variables.

Most notably, the total joint entropy \(H(S,P)\) often interpreted as a measure of global phase-space delocalization also peaks in the vicinity of the avoided crossing (see inset of Fig.~\ref{Figure-4}c). Importantly, this increase cannot be attributed solely to independent broadening of the marginal distributions. Rather, it reflects a non-negligible contribution from mutual information. In other words, the rise in total entropy is driven not just by spatial dispersion but also by the emergence of structured inter-variable dependence. This result highlights a subtle but crucial distinction: even within a strongly chaotic regime, enhanced coherence can coexist with delocalization-a counterintuitive, yet revealing, signature of order emerging from disorder.

\section{Universality Across Chaotic Systems}
To assess the generality of our findings, we apply the same informational framework to a geometrically distinct system: the oval billiard, simulated with parameters \(a = 1\), \(b = 1.03\), and a deformation parameter \(\varepsilon\) scanned in steps of \(\Delta\varepsilon = 1 \times 10^{-5}\). All computational steps ranging from numerical solution of the Helmholtz equation to entropy based analysis are identical to those used in the quadrupole case.

Figure~\ref{Figure-5}(a) displays eigenvalue trajectories as a function of \(\varepsilon\), revealing a well-defined avoided crossing between two branches. Panel (b) presents the real-space intensity plots for representative eigenmodes A-F, along with the corresponding configuration-space mutual information \(I(X;Y)\). Panel (c) shows the phase-space mutual information \(I(S;P)\) for two pairs of eigenmodes. As in the quadrupole case, \(I(S;P)\) exhibits a pronounced peak near the avoided crossing, indicating enhanced statistical dependence between conjugate phase-space variables.

Despite the differences in boundary geometry and deformation parameterization, the overall informational behavior remains consistent: configuration-space mutual information \(I(X;Y)\) varies smoothly without a distinct peak, whereas \(I(S;P)\) in phase space shows a clear and localized maximum near the crossing. These results confirm that phase-space mutual information serves as a robust, geometry independent marker of eigenmode hybridization and internal restructuring in chaotic quantum systems.

\section{Classical Invariant Tori and Multi-Branch Functionality}
Classical invariant tori in the two-dimensional phase space (Poincare surface of section, or PSOS) are typically one-dimensional manifolds that need not be single-valued functions of \(s\); they may form closed loops or multiply connected curves. Nevertheless, \emph{locally} such structures can be approximated by a relation of the form \(p \approx f(s)\), indicating a near-deterministic mapping from \(s\) to \(p\). If the Husimi distribution of a quantum eigenmode is concentrated around such a manifold with Gaussian width \(\sigma\), then the phase-space density takes the approximate form
\begin{align}
h(s, p) \propto \exp\!\biggl[-\,\frac{(p - f(s))^2}{2\,\sigma^2}\biggr],
\end{align}
which yields a low conditional entropy \(H(P \mid S)\), and hence a high mutual information \(I(S;P)\). In the idealized limit \(\sigma \to 0\), the mutual information approaches \(\min\{H(S), H(P)\}\), reflecting an (almost) invertible functional dependence between conjugate variables.

During an avoided crossing, multiple such tori may be simultaneously populated, each associated with a distinct classical orbit family. In this case, the total support of the Husimi distribution in phase space becomes broader, increasing the total entropy \(H(S,P)\). Yet, within each locally supported region, the distribution remains close to a function-like form. This allows for strong local correlations to coexist with global delocalization-an outcome that may appear counterintuitive, but is physically well-motivated. Consequently, both \(H(S,P)\) and \(I(S;P)\) can increase together, challenging the conventional expectation that increasing chaoticity suppresses coherent internal structure.

\begin{table*}[t]
\caption{\label{tab:modeComparison}%
Comparison of conditional entropies and mutual information for {\bf Mode~1} and {\bf Mode~2} at the center of the avoided crossing. Both exhibit an entropy peak, but differ in correlation strength.}
\centering
\begin{ruledtabular}
\begin{tabular}{llcc}
\textbf{Quantity} & \textbf{Functional structure} & \textbf{Mode 1} & \textbf{Mode 2} \\
\hline
$H(S|P)$ & $s \approx g(p)$ & increasing & decreasing \\
$H(P|S)$ & $p \approx f(s)$ & increasing & increasing \\
$H(S,P)$ & delocalization & increasing & increasing \\
$I(S;P)$ & $p\approx g^{-1}$ & weak & stronger
\end{tabular}
\end{ruledtabular}
\end{table*}

To quantify the coherence asymmetry observed in phase space, we define a normalized invertibility measure:
\begin{equation}
\textsc{Inv}_1 = \frac{I(S;P)}{\min\{H(S), H(P)\}},
\end{equation}
which ranges from 0 (statistical independence) to 1 (perfect bijection). While mutual information $I(S;P)$ quantifies overall correlation between conjugate variables, it does not distinguish between diffuse statistical dependence and near-deterministic functional structure. The metric $\textsc{Inv}_1$ instead captures how well one variable constrains the other, providing a quantitative proxy for partial invertibility.
As summarized in Table~\ref{tab:modeComparison}, {\bf Mode~2} exhibits a sharper rise in $I(S;P)$ than {\bf Mode~1}, despite both showing comparable increases in total entropy $H(S,P)$. This difference reflects a more function-like alignment between $s$ and $p$, consistent with the observed directional entropy asymmetry. In our simulations, $\textsc{Inv}_1$ closely tracks $I(S;P)$ in shape (with reduced magnitude), indicating that partial invertibility strengthens with phase-space delocalization. This suggests that, contrary to expectation, coherence can persist and even emerge under increasing chaos.

\section{Conclusion}
We have demonstrated that in chaotic quantum billiards, avoided crossings can enhance coherence between conjugate variables in phase space. Rather than simply mixing eigenstates, these transitions increase mutual information, providing a distinct entropic signature of internal structure. This challenges the conventional assumption that increasing chaoticity necessarily suppresses coherence, and instead points to a robust, information-theoretic manifestation of quantum-``classical correspondence.

Our findings reveal that delocalization and coherence are not mutually exclusive, but can emerge simultaneously as complementary features of chaotic eigenstates. This duality broadens our understanding of how structure can persist or even be amplified ''within disordered quantum systems. More broadly, our results suggest that phase-space coherence, as quantified by mutual information, may serve as a universal diagnostic of internal organization in wave-based chaotic systems beyond billiards, including optical, acoustical, and matter-wave platforms.

\section{acknowledgement}
This work was supported by the National Research Foundation of Korea (NRF) through a grant funded by the Ministry of Science and ICT (Grant No. RS-2023-00211817 and No. RS-2025-00515537), the Institute for Information \& Communications Technology Promotion (IITP) grant funded by the Korean government (MSIP) (Grant No. RS-2019-II190003 and No. RS-2025-02304540), and Korea Institute of Science and Technology Information (KISTI).


\end{document}